\documentclass[12pt,prl,aps,superscriptaddress]{revtex4}
\usepackage{graphicx}
\usepackage{amssymb,bm}
\newcommand{\beq}[2]{\begin{equation}#1\label{#2}\end{equation}}
\newcommand{\ceq}[1]{(\ref{#1})}

\newcommand{\intsym}{\int_0^{t_f}dt\int_0^Lds}

\newfont{\mbld}{cmbx10 scaled 800}
\newfont{\cab}{cmsy10 scaled 1200}
\newfont{\scab}{cmsy10 scaled 1000}
\newfont{\bcall}{cmbsy10 scaled 1200}

\begin{document}
\title{On the connection of the generalized nonlinear sigma model with
constrained stochastic dynamics}
\author{Franco Ferrari}
\email{ferrari@univ.szczecin.pl}
\author{Jaros{\l}aw Paturej}\email{jpaturej@univ.szczecin.pl}
\affiliation{Institute of Physics and CASA*, University of Szczecin,
  ul. Wielkopolska 15, 70-451 Szczecin, Poland}

\begin{abstract}
The dynamics of a freely jointed chain in the continuous limit is
described by a field theory which closely resembles  the
nonlinear sigma model. The generating functional $\Psi[J]$ 
of this field theory  contains nonholonomic
constraints, which are imposed by inserting in the
path integral expressing $\Psi[J]$ a suitable product of
delta functions. The same procedure 
is commonly applied in statistical mechanics in order to enforce
topological conditions on a system of linked polymers.
The disadvantage of this method is that the contact with the stochastic
process governing the diffusion of the  
chain is apparently lost.
The main goal of this work is to reestablish this contact. 
To this purpose, it is shown here that the generating
functional $\Psi[J]$ coincides with the generating functional of the
correlation functions of the solutions of a constrained
Langevin equation. In the discrete case, this Langevin equation
describes as expected
the Brownian motion of beads connected together by links of
fixed length.
\end{abstract}
\maketitle
\section{Introduction}\label{sec:intro}
The subject of this work is a chain obtained by performing the
continuous limit of a system of $N-1$ links of fixed length $a$ and
$N$ beads of constant mass $m$. In this limit the number $N$ of beads
approaches infinity, the length of the links and the mass of the beads
go to zero, while the total length $L$ of the chain remains finite.
The dynamics of a
chain with rigid constraints of this kind has been studied in a
remarkable series of papers \cite{EdwGoo,EdwGoo2,EdwGoo3} using an
approach based on the Langevin equation. Later on, mainly the
statistical mechanics of such chains has been investigated, see e.~g.
\cite{grosberg,mazars, a-e}. Dynamical models are however
interesting by themselves and have also some
 applications, for instance in modeling 
the response of a chain to mechanical stresses in 
micromanipulations \cite{bustamante}.

In Ref.~\cite{FEPAVI1} 
the dynamics of the constrained chain
has been considered using path integral methods.
The resulting model  is
a generalization of the nonlinear sigma model \cite{nlsigma} which
will be called here
the generalized 
nonlinear sigma model or simply GNL$\sigma$M.
The most striking difference from the standard nonlinear sigma model
is that in the GNL$\sigma$M the constraint is nonholonomic. 
The relation of the
GNL$\sigma$M with the Rouse model \cite{rouse,doiedwards} has been
discussed in Ref.~ \cite{FEPAVI1}. It has also been shown that it
gives the correct 
equilibrium limit in agreement with
Ref.~\cite{EdwGoo}.
Applications of the GNL$\sigma$M have been developed in
Refs.~\cite{FEPAVI2,FEPAVI3}, computing 
for instance the dynamic form factor of the
chain in the semiclassical approximation and
the probability distribution
$Z(\mathbf r_{12})$ which measures the probability that in a given
interval of time the average distance between two points of
the chain is $\mathbf r_{12}$.

One point that still needs to be clarified is if the GNL$\sigma$M can
be related  to 
 some stochastic process. 
In fact, the GNL$\sigma$M has not been derived starting from a Langevin
equation and applying for instance the Martin--Siggia--Rose formalism
\cite{MSR} in order to pass  the path integral formulation.
The problem is that this approach becomes cumbersome if one has to
deal with constraints.
For this reason, in \cite{FEPAVI1}
the constraints have been added
to the path integral describing the
fluctuations
of the beads  with the help of an insertion of Dirac delta functions.
This is a widely exploited procedure   in the statistical
mechanics of polymers in order to impose topological conditions
\cite{edwa,vilbre, FeKlLa, kleinertpi}.

To establish a relation between the GNL$\sigma$M and a stochastic
process
is the main goal of the present work.
To this purpose, after a brief introduction to the GNL$\sigma$M, we
define in the next Section a two dimensional vector field
$\bm\varphi_{\bm\nu}$ which
satisfies a free Langevin equation and additional nonholonomic
constraints. These are exactly the same constraints which appear also in the
GLN$\sigma$M. Our treatment is limited to two dimensions for
simplicity.
The generating functional $\tilde\Psi[J]$ 
of the
correlation functions of the fields $\bm\varphi_{\bm\nu}$ can be
constructed using the prescription of \cite{zinn}.
The discretized version of $\tilde\Psi[J]$ describes the Brownian
motion of a set of $N-$beads with diffusion constant $D$
which are connected together by links of fixed length.
The difference between  $\tilde\Psi[J]$ 
and
the generating functional  $\Psi[J]$  of the correlation
functions of the GNL$\sigma$M consists in a functional determinant.
We show that this determinant is trivial  by
eliminating the constraints using a special set of
variables, called here pseudo--polar coordinates.
As a result we prove the equivalence of $\tilde\Psi[J]$  and
$\Psi[J]$ and thus the connection of the GNL$\sigma$M with a
stochastic process of diffusing particles.
\section{The generalized nonlinear sigma model and its relation to the
Langevin equation}
The starting point of this Section is the generating functional of the
GNLSM presented in Ref.~\cite{FEPAVI1}:
\beq{
\Psi[J]=\int{\cal D}\mathbf R(t,s) e^{-
  c\int_0^{t_f}dt\int_0^Nds\dot{\mathbf R}^2(t,s)}
\delta(|\mathbf R'(t,s)|^2-1) e^{-\int_0^{t_f}dt\int_0^Lds\mathbf
  J(t,s)\cdot \mathbf R(t,s)}
}{genfunzero}
with $\dot\mathbf R=\partial \mathbf R/\partial t$ and
$\mathbf R'=\partial \mathbf R/\partial s$.
The boundary conditions at $t=0$ and $t=t_f$ of the field $\mathbf R(t,s)$
are respectively given by $
\mathbf R(0,s)=\mathbf R_0(s)$
and $\mathbf
R(t_f,s)=\mathbf R_f(s)$, where
$\mathbf R_0(s)$ and $\mathbf R_f(s)$ represent given static
conformations of the chain.
For a
ring--shaped chain the boundary conditions with respect to $s$ are
periodic: 
$
\mathbf R(t,s)=\mathbf R(t,s+L)$. An open chain with fixed ends
 may be described using instead the boundary conditions: $\mathbf
 R(t,0)=\mathbf r_1$, $\mathbf R(t,L)=\mathbf r_2$, $\mathbf r_1$ and
 $\mathbf r_2$ being the locations of the fixed ends. 
It was shown in Refs. \cite{FEPAVI1} and \cite{FEPAVI2} that the above
generating functional describes the dynamics of a closed chain that is
the continuous version of a freely jointed chain consisting of links
and beads. 
The constant $c$ appearing in Eq.~\ceq{genfunzero} is
given by:
\beq{
c=\frac{M}{4k_BT\tau L}
}{cdef}
Here $k_B$ denotes the Boltzmann constant, $T$ is the temperature and
$\tau$ is the relaxation time which characterizes the ratio of the
decay of the drift velocity of the beads.
$M$ and $L$ represent the total mass and the total length of the chain
respectively.
The starting point to  derive $\Psi[J]$ is the path integral $\Psi_N$
describing the brownian motion of a set of $N$ particles.
The rigid constraints, which take into account the fact that these
particles form a chain and thus are connected together by $N-1$
massless segments of fixed length $a$, are introduced in the path integral
with the help of a suitable product of Dirac delta functions.
The limit of $\Psi_N$ from the discrete system to the continuous chain
in which $N\longrightarrow +\infty$, $a\longrightarrow 0$ and $Na=L$ can be 
performed rigorously. The result is the generating functional of
Eq.~\ceq{genfunzero}.
This procedure is different from the usual approach to the dynamics of
a chain, which is based on a Langevin equation. In this Section we
are going to show that the GNL$\sigma$M can be related to a Langevin
equation too. 
For simplicity, we restrict ourselves to the two dimensional case.

Since the GNL$\sigma$M ignores all interactions, it is natural to
suppose that it should be related to a Langevin equation with no
external forces:
\beq{
\dot{\bm \varphi}_{\bm\nu}=\bm\nu
}{langevequ}
where $\bm\varphi_{\bm \nu}$ is a two dimensional vector field and
$\bm\nu$ is a white noise source, whose components $\nu^{(i)}$, $i=1,2$
satisfy the basic correlation functions:
\beq{\langle
\nu^{(i)}(t,s))
\rangle=0}{nuione}
\beq{\langle
\nu^{(i)}(t,s)\nu^{(j)}(t',s')
\rangle=\frac{\delta^{ij}}{c}\delta(t-t')\delta(s-s')
\qquad\qquad i,j=1,2
}{nuinujtwo}
One may also expect that, together with Eq.~\ceq{langevequ}, the field
$\bm\varphi_{\bm \nu}$ must also satisfy the constraint:
\beq{
\bm\varphi_{\bm\nu}^{\prime\,2}=1
}{constrvarphinu}
The generating functional $\tilde\Psi[J]$ of the correlation functions of
the field $\bm\varphi_{\bm\nu}$ is then given by \cite{zinn}:
\beq{
\tilde\Psi[J]=\int_{\bm\varphi_{\bm\nu}^{\prime\,2}=1}{\cal D}\bm\nu
e^{-c\intsym\bm\nu^2} e^{\intsym \mathbf J\cdot\bm\varphi_{\bm\nu}}
}{psitildej}
The meaning of the statistical sum in the right hand side of the above
equation becomes clear if we rewrite it as follows:
\beq{
\tilde\Psi[J]=\int{\cal D}\bm\nu\int_{\mathbf R^{\prime\, 2}=1}{\cal
  D}\mathbf R e^{-c\intsym\bm\nu^2} 
\delta(\mathbf R -\bm\varphi_{\bm\nu})
e^{\intsym \mathbf J\cdot
  \mathbf R}
}{psitildejel} 
The path integration over $\bm\nu$ is now unconstrained, while that
over the new field $\mathbf R$ is limited to the configurations which
are of the form:
\beq{\mathbf
  R(t,s)=\int_0^sdu(\cos\phi(t,u),\sin\phi(t,u))+\mathbf
  R_0(t)}{integconfigs}  
where $\mathbf R_0(t)$ is independent of $s$. The only left degree of
freedom is the angle $\phi(t,s)$.

The generating functional $\Psi[J]$ of Eq.~\ceq{genfunzero}
differs from $\tilde\Psi[J]$  due to the presence of the functional
Dirac delta function $\delta(\mathbf R'^2-1)$. As a matter of fact, it
is easy to show that:
\beq{
\Psi[J]=\int{\cal D}\bm \nu\int{\cal D}\mathbf R 
 e^{-c\intsym \bm
  \nu^2}\delta(\mathbf R'^2-1)
\delta(\dot{\mathbf R}-\bm\nu)
e^{-\intsym \mathbf J\cdot\mathbf R}
}{genfunzeronewform}
The connection with the Langevin equation \ceq{langevequ} is made by
noticing that, for any solution $\bm\varphi_{\bm\nu}$ of that equation
it is possible to write  the formula:
\beq{
\delta(\dot{\mathbf R}-\bm\nu)={\det}^{-1}\partial_t\delta(\mathbf
  R-\bm\varphi_{\bm\nu})
}{diracdeltasubs}

Applying Eq.~\ceq{diracdeltasubs} to Eq.~\ceq{genfunzeronewform}
we obtain, up to an irrelevant constant:
\beq{\Psi[J]=\int{\cal D}\mathbf R{\cal D}\bm \nu
e^{-c\intsym \bm
  \nu^2}
\delta(\mathbf R-\bm\varphi_{\bm \nu})\delta(\mathbf R'^2-1)
e^{-\intsym \mathbf J\cdot\mathbf R}
}{beforedelta}
As already announced, this expression of the generating functional
$\Psi[J]$ differs 
from
$\tilde\Psi[J]$ only by the fact that the condition $\mathbf R'^2=1 $
is imposed with the help of the delta function $\delta(\mathbf
R'^2-1)$.
In the next Sections the degrees of freedom which are
frozen by the condition
$\mathbf R'^2=1$ will be projected out from the path integral
\ceq{beforedelta} and it will be shown that what remains is exactly
the generating functional $\tilde\Psi[J]$  related to the
constrained stochastic process of Eqs.~\ceq{langevequ} and
\ceq{constrvarphinu}.
\section{The discrete generating functional in pseudo--polar
  coordinates }
As a first step to show the equivalence of the generating functionals
$\Psi[J]$ and $\tilde\Psi[J]$ we replace the continuous variables $s$
and $t$ with discrete variables $s_m$ and $t_n$, with $0\le m\le M$
and $0\le n\le N$. The spacings in the discrete $s$ and $t-$lines are
respectively given by:
\begin{eqnarray}
s_m-s_{m-1}&=&a\qquad\qquad m=2,\ldots,M\\
t_n-t_{n-1}&=&b\qquad\qquad n=2,\ldots,N
\end{eqnarray}
where $a$ and $b$ are supposed to be very small.
The continuous limit is recovered in the limit $M,N\longrightarrow
+\infty$, $a,b\longrightarrow 0$ and $Ma=L$, $Nb=t_f$.
To simplify formulas, it will be used in the following the shorthand
notation: 
\beq{
\mathbf R(t_n,s_m)\equiv \mathbf R_{nm}\qquad\qquad
\bm \nu(t_n,s_m)\equiv \bm \nu_{nm}\qquad\qquad
\bm \varphi_{\bm\nu}(t_n,s_m)\equiv \bm \varphi_{\bm\nu,nm}
}{shorthandnot}
In this way the discrete version of the constraint $\mathbf
R'^2(t,s)=1$ is replaced by the 
set of conditions:
\beq{
\frac{
(\mathbf R_{nm}-\mathbf R_{n(m-1)})
}{a^2}=1\qquad\qquad\begin{array}{c}
n=1,\ldots,N\\
m=2,\ldots,M
\end{array}
}{constao}
With the above settings the generating functional $\Psi[J]$ of
Eq.~\ceq{beforedelta} may be rewritten as follows~\footnote{Unless
  otherwise specified, from now on it will understood that the 
indices $n$ and $m$ in
sums and products will take all possible values in their respective ranges,
i. e. $1\le n\le N$ and $1\le m\le M$.}:
\begin{eqnarray}
\Psi[J]&=&\lim_{N\to\infty}\lim_{M\to\infty}\int_{-\infty}^{+\infty}\left[
\prod_{n,m}d\bm\nu_{nm} 
d\mathbf R_{nm}\right]\exp\left\{
-abc\sum_{n,m}\bm\nu^2_{nm}
\right\}\nonumber\\
&\times&\prod_{n,m}\delta(\mathbf R_{nm}-\bm\varphi_{\bm\nu,nm})
\exp\left\{
ab\sum_{n,m}\mathbf J_{nm}\mathbf R_{nm}
\right\}\nonumber\\
&\times&\prod_n\prod_{m=2}^M\frac2a\delta\left(\frac{|\mathbf R_{nm}-\mathbf
R_{n(m-1)}|}{a^2}-1\right)\label{psijdiscr} 
\end{eqnarray}
Let us also note in the last line of equation \ceq{psijdiscr} the
normalization 
factor  $\prod_n\prod_{m=2}^M\frac2a$ in 
the definition of the delta function imposing the constraints. The
reason of this normalization will be clear later.
Without the constraints, the above equation would describe a discrete
chain of $N-1$ segments of 
length $a$ and
$N$ beads of mass $m$ respectively which perform a Brownian motion.
The diffusion constant $D$ is recovered from the parameter $c$ of
Eq.~\ceq{cdef} as follows. First of all, we note that
$ca=\frac1{4k_BT\tau}\frac M{L}a$. The ratio $\frac ML$ is nothing
but the linear density 
of mass along the chain, so that $\frac ML a$ is equal to the mass $m$ of a
single bead, i.~e.: $\frac ML a=m$. As a consequence, $ca=\frac
m{4k_BT\tau}$.
At this point we remember that the mobility of a particle $\mu$  may be
expressed in terms of the particle mass $m$ and of the relaxation time
$\tau$ as follows: $\frac m\tau=\frac 1\mu$. Thus, $ca=\frac
1{4k_BT\mu}$.
Due to the fact that $D=k_BT\mu$, it is possible to write $ca=\frac
1{4D}$.

To eliminate the constraints \ceq{constao}, we pass to a new set of
coordinates, which in the following will be called pseudo--polar:
\beq{
\mathbf R_{nm}=\sum_{m'=1}^Ml_{nm'}\left(
\cos\phi_{nm'},\sin\phi_{nm'}
\right)
}{pseudopolar}
The ranges of variation of the variables $l_{mn}$ and $\phi_{nm}$ are
respectively given by:
\beq{
0\le l_{mn}<+\infty\qquad\qquad 0\le\phi_{nm}\le 2\pi
}{rangevarppc}
The coordinate $l_{nm}$ for $n=1,\ldots,N$ and $m=2,\ldots,M$, describes the
length of the $m-$th segment at the instant $t_n$. The coordinate
$l_{n1}$ is very special, because it
gives the position with respect to the origin of the reference system
of the first bead in the chain at the time $t_n$.
Finally, the angles $\phi_{nm}$ tell us how the $N-1$ segments are
reciprocally oriented.
After
the transformation
\ceq{pseudopolar}, the vector $\mathbf R_{nm}$ depends on the
variables $l_{nm}$  and $\phi_{nm}$, i. e.:
\beq{
\mathbf R_{nm}=
\mathbf R_{nm}(\{l_{nm}\},l_{n1};\{\phi_{nm}\})}{restranfs}
where $\{l_{nm}\}$ is the set of all $l_{nm}$'s for which
$m\ne 2$ and $\{\phi_{nm}\}$ is the set of all $\phi_{nm}$'s.
Analogously, we denote with $\{\mathbf R_{nm}\}$ the set of all
$\mathbf R_{nm}$'s for $m=1,\ldots,M$ and $n=1,\ldots,N$.
We are now able to explain the reason of the normalization factor
 $\prod_n\prod_{m=2}^M\frac2a$ in Eq.~\ceq{psijdiscr}.
In the pseudo-polar variables the constraints \ceq{constao} become:
$\frac {l_{nm}^2}{a^2}=1$. The factor $\frac 2a$ is necessary in order
to normalize the delta functions imposing these constraints.
As a matter of fact, it is possible to check that
 $\frac 2a\int_0^{+\infty}dl_{nm}\delta(\frac{l^2_{nm}}{a^2}-1)=1$.

In order to perform the transformations \ceq{pseudopolar} in the
expression of the generating functional $\Psi[J]$ of
Eq.~\ceq{psijdiscr}, we need to compute the associated Jacobian determinant.
In the rest of this Section we will prove 
for a general functional $f(\{\mathbf R_{nm}\})$
the following formula:
\beq{
\int\prod_{n,m}d\mathbf  R_{nm}f(\{\mathbf R_{nm}\}) =
\int_0^{+\infty}\prod_{n,m} dl_{nm} \int_0^{2\pi}\prod_{n,m} d\phi_{nm}
f(\{R_{nm}(\{l_{nm}\},l_{n1};\{\phi_{nm}\})\})J_{NM}
}{fdfdf}
where   the Jacobian $J_{NM}$  of the transformation \ceq{pseudopolar}
is given by:
\beq{
J_{NM}(\{l_{nm}\},l_{n1};\{\phi_{nm}\})=\prod_n l_{nM}l_{n(M-1)}\cdots l_{n1}
}{jacobian}

Let's show that $J_{NM}$ is really that given in
Eq.~\ceq{jacobian}. In order to proceed, it is convenient 
to introduce the components $x^{(1)}_{nm}$ and $x^{(2)}_{nm}$ of the
vectors $\mathbf R_{nm}$, i.~e. $\mathbf
  R_{nm}=(x^{(1)}_{nm},x^{(2)}_{nm})$. Thus, Eq.~\ceq{pseudopolar}
    becomes:
\beq{
x^{(1)}_{nm}=\sum_{m'=1}^ml_{nm'}\cos\phi_{nm'}\qquad\qquad 
x^{(2)}_{nm}=\sum_{m'=1}^ml_{nm'}\sin\phi_{nm'}\qquad\qquad 
}{pseudopolarII}
and $J_{NM}$ may be written as follows:
\beq{
J_{NM}(\{l_{nm}\},l_{n1};\{\phi_{nm}\})=\det\left|
\begin{array}{cc}
\frac{\partial x^{(1)}_{nm} }
{\partial l_{n'm'}}
& \frac{\partial x^{(2)}_{nm} }
{\partial l_{n'm'}}\\
\frac{\partial x^{(1)}_{nm} }{\partial \phi_{n'm'}}
& \frac{\partial x^{(2)}_{nm} }{\partial \phi_{n'm'}}
\end{array}
\right|
}{exprJNM}

Strictly speaking, $J_{NM}$
is the determinant of a block matrix
$A_{nm;n'm'}$ with composite indices $nm$ and ${n'm'}$.
$A_{nm;n'm'}$ is composed by four $NM\times NM$ matrices, since
$n,n'=1,\ldots,N$ and $m,m'=1,\ldots,M$.
Due to the fact that
$\frac{\partial x^{(i)}_{nm} }
{\partial l_{n'm'}}=\frac{\partial x^{(i)}_{nm} }
{\partial \phi_{n'm'}}=0$
for $i=1,2$  if $n\ne n'$, $A_{nm;n'm'}$ is a block diagonal matrix.
As a consequence, it is possible to write its determinant as follows:
\beq{
J_{NM}=\prod_nJ_{nM}
}{jnmjNm}
where
\beq{
J_{nM}=
\det\left|
\begin{array}{cc}
\frac{\partial x^{(1)}_{nm} }
{\partial l_{nm'}}
& \frac{\partial x^{(2)}_{nm} }
{\partial l_{nm'}}\\
\frac{\partial x^{(1)}_{nm} }{\partial \phi_{nm'}}
& \frac{\partial x^{(2)}_{nm} }{\partial \phi_{nm'}}
\end{array}
\right|
}{jnMdef}
Using Eqs.~\ceq{pseudopolarII}, one finds
after a few calculations 
that $J_{nM}$ is the determinant of the block matrix:
\beq{
J_{nM}=\det\left|
\begin{array}{cc}
A(n)&B(n)\\
C(n)&D(n)
\end{array}
\right|
}{jnmdef}
$A(n),B(n),C(n),D(n)$ are lower triangular $M\times M$ matrices with
elements: 
\begin{eqnarray}
A_{mm'}(n)=\theta_{mm'}\cos\phi_{nm'}&\qquad\qquad
&B_{mm'}(n)=\theta_{mm'}\sin\phi_{nm'} \\
C_{mm'}(n)=-l_{nm'}\theta_{mm'}\sin\phi_{nm'}&\qquad\qquad
&D_{mm'}(n)=l_{nm'}\theta_{mm'}\cos\phi_{nm'} 
\end{eqnarray}
Here the matrix $\theta_{mm'}$ denotes the discrete equivalent of the
Heaviside theta-function:
\begin{eqnarray}
\theta_{mm'}=1&\enskip\mbox{if}\enskip& m'\le m\\
\theta_{mm'}=0&\enskip\mbox{if}\enskip& m'> m
\end{eqnarray}
If the matrices $A(n),B(n),C(n),D(n)$ would commute, one could use a
known theorem of linear algebra and write:
$J_{nM}=\det(A(n)D(n)-B(n)C(n))$.
In our case these matrices do not commute, but it is still possible to
compute the determinant $J_{nM}$ by induction on $M$.

If $M=1$ it is easy to show that:
\beq{J_{n1}=l_{n1}}{inductionfirst}
Next, we prove that
\beq{
J_{nM}=l_{nM}J_{n(M-1)}
}{reccentral}
To this purpose, it will be convenient to introduce new indices
$\alpha,\beta=1,\ldots,M-1$.
At this point, we note that the $M-$th column of the $2M\times 2M$
block matrix whose 
determinant we wish to compute in Eq.~\ceq{jnmdef} has only two
elements which are not zero. Thus, we expand $J_{nM}$ with respect to
the $M-$th column. Taking into account the necessary permutations and
the fact that the two  nonvanishing elements are $A_{MM}(n)=\cos\phi_{nM}$
and $C_{MM}(n)=-l_{nm}\sin\phi_{nM}$ we obtain:
\begin{eqnarray}
J_{nM}&=&\cos\phi_{nM}\det\left|
\begin{array}{ccc}
\theta_{\alpha\beta}\cos\phi_{n\beta}&
\theta_{\alpha\beta}\sin\phi_{n\beta}&0\\
-l_{n\beta}\theta_{\alpha\beta}\sin\phi_{n\beta}&
l_{n\beta}\theta_{\alpha\beta}\cos\phi_{n\beta}&0\\
-l_{n\beta}\theta_{M\beta}\sin\phi_{n\beta}&
l_{n\beta}\theta_{M\beta}\cos\phi_{n\beta}&
l_{nM}\cos\phi_{nM}
\end{array}
\right|\nonumber\\
&+&(-1)^Ml_{nM}\sin\phi_{nM}\det
\left|
\begin{array}{ccc}
\theta_{\alpha\beta}\cos\phi_{n\beta}&
\theta_{\alpha\beta}\sin\phi_{n\beta}&0\\
\theta_{M\beta}\cos\phi_{n\beta}&
\theta_{M\beta}\sin\phi_{n\beta}&\sin\phi_{nM}\\
-l_{n\beta}\theta_{\alpha\beta}\sin\phi_{n\beta}&
l_{n\beta}\theta_{\alpha\beta}\cos\phi_{n\beta}&
0
\end{array}
\right|
\end{eqnarray}
The  determinants of the remaining two $(2M-1)\times(2M-1)$ matrices
may be expanded according to the $(2M-1)-$th column, because these columns
contain only one nonvanishing element.
After simple calculations one finds:
\beq{
J_{nM}=l_{nM}\det\left|
\begin{array}{cc}
\theta_{\alpha\beta}\cos\phi_{n\beta}&\theta_{\alpha\beta}\sin\phi_{n\beta}\\
-l_{n\beta}\theta_{\alpha\beta}\sin\phi_{n\beta}&l_{n\beta}\theta_{\alpha\beta}
\cos\phi_{n\beta}
\end{array}
\right|
}{finalproof}
which is exactly Eq.~\ceq{reccentral} because
\beq{
J_{n(M-1)}=\det\left|
\begin{array}{cc}
\theta_{\alpha\beta}\cos\phi_{n\beta}&\theta_{\alpha\beta}\sin\phi_{n\beta}\\
-l_{n\beta}\theta_{\alpha\beta}\sin\phi_{n\beta}&l_{n\beta}\theta_{\alpha\beta}
\cos\phi_{n\beta}
\end{array}
\right|
}{fdfsfd}
Using Eqs.~\ceq{inductionfirst} and \ceq{reccentral} it is easy to
show by induction that $J_{nM}=l_{nM}l_{n(M-1)}\cdots l_{n1}$.
With a straightforward application of Eq.~\ceq{jnmjNm} it is now
possible to prove
Eq.~\ceq{jacobian}.
\section{Recovering the generating functional $\tilde\Psi[J]$ of the
 constrained stochastic process of
 Eqs.~\ceq{langevequ}--\ceq{constrvarphinu} }
Let's now go back to the generating functional
$\Psi[J]$ of Eq.~\ceq{psijdiscr}.
After the change of variables  \ceq{pseudopolar}, the delta functions
imposing the constraints simplify as follows: $\delta\left(\frac{|\mathbf
  R_{nm}-\mathbf 
R_{n(m-1)}|}{a^2}-1\right)=\delta\left(
\frac{l_{nm}^2}{a^2}-1
\right)$. Further simplifications are obtained after applying the two delta
function identities 
$\delta\left(
\frac{l_{nm}^2}{a^2}-1
\right)=a^{2}\delta(l_{nm}^2-a^2)$ and
$\delta(l_{nm}^2-a^2)=\frac 1{2a}\left[
\delta(l_{nm}-a)+\delta(l_{nm}+a)
\right]$. Remembering that in our case $l_{nm}\ge 0$, it is
possible to put: $\delta(l_{nm}^2-a^2)=\frac 1{2a}
\delta(l_{nm}-a)$.
As a consequence, the expression
of the generating functional $\Psi[J]$ in pseudo--polar coordinates becomes:
\begin{eqnarray}
\Psi[J]&=&\lim_{N\to\infty}\lim_{M\to\infty}\int_{-\infty}^{+\infty}
\prod_{n,m}d\bm\nu_{nm} 
\int_0^{+\infty}\prod_{n,m}dl_{nm}\int_0^{2\pi}\prod_{n,m}d\phi_{nm}
\exp\left\{
-abc\sum_{n,m}\bm\nu^2_{nm}
\right\}\nonumber\\
&\times&\prod_{n,m}\delta(\mathbf R_{nm}(\{l_{nm}\},l_{n1};\{\phi_{nm}\})
-\bm\varphi_{\bm\nu,nm})
\exp\left\{
ab\sum_{n,m}\mathbf J_{nm}\cdot\mathbf R_{nm}(\{l_{nm}\},l_{n1};\{\phi_{nm}\})
\right\}\nonumber\\
&\times&
\prod_nl_{n1}
\left[\prod_n\prod_{m=2}^M\delta\left(l_{nm}-a\right)\right]
\prod_{m=2}^Ml_{nM}\cdots l_{n2}
\label{psijdiscr1} 
\end{eqnarray}
In writing the above equation we have separated from the Jacobian
determinant $J_{NM}$ the contribution coming from the $l_{n1}'s$,
because these quantities denote the positions 
 with respect to the origin
of the first bead at
different times $t_n$'s and are thus not fixed by the constraints.
The integration in Eq.~\ceq{psijdiscr1} over the $l_{nm}$'s, for
$n=1,\ldots,N$ and $m=2,\ldots,M$, 
produces as a result:
\begin{eqnarray}
\Psi[J]&=&\lim_{N\to\infty}\lim_{M\to\infty}
a^{N(M-1)}
\int_{-\infty}^{+\infty}
\prod_{n,m}d\bm\nu_{nm} 
\int_0^{+\infty}\prod_{n}dl_{n1}l_{n1}\int_0^{2\pi}\prod_{n,m}d\phi_{nm}
\exp\left\{
-abc\sum_{n,m}\bm\nu^2_{nm}
\right\}\nonumber\\
&\times&\prod_{n,m}\delta(\mathbf R_{nm}(\{a\},l_{n1};\{\phi_{nm}\})
-\bm\varphi_{\bm\nu,nm})
\exp\left\{
ab\sum_{n,m}\mathbf J_{nm}\cdot\mathbf R_{nm}(\{a\},l_{n1};\{\phi_{nm}\})
\right\}\label{psijdiscr2} 
\end{eqnarray}
Here the symbol $\{a\}$ denotes the set of all $l_{nm}$'s for $m\ne
2$ after the imposition of the constraints $l_{nm}=a$.
We can now rewrite Eq.~\ceq{psijdiscr2} as an integral over a
restricted domain $D$:
\begin{eqnarray}
\Psi[J]&=&\lim_{N\to\infty}\lim_{M\to\infty}\int_{D}\prod_{n,m}dl_{nm}d\phi_{nm}
\int_{-\infty}^{+\infty}\prod_{n,m}d\bm\nu_{nm}\exp\left\{
-abc\sum_{n,m}\bm\nu^2_{nm}
\right\}
\prod_n l_{nM}\cdots l_{n1}
\nonumber\\
&\!\!\!\!\!\!\!\!\!\!\!\!\!\!\!\!\!\!\!\!\!\!\!\!\!
\times&
\!\!\!\!\!\!\!\!\!\!\!\!\!\!\!
\prod_{n,m}\delta(\mathbf R_{nm}(\{l_{nm}\},l_{n1};\{\phi_{nm}\})
-\bm\varphi_{\bm\nu,nm})\exp\left\{
ab\sum_{n,m}\mathbf J_{nm}\cdot\mathbf R_{nm}(\{l_{nm}\},l_{n1};\{\phi_{nm}\})
\right\} 
\end{eqnarray}
where $D$ is the domain of all $l_{nm}$'s and $\phi_{nm}$'s with the
constraints $l_{nm}=a$ for $m=2,\ldots,M$ and $n=1,\ldots,N$:
\beq{
D=\left\{
\{l_{nm}\},\{\phi_{nm}\}\left|
\begin{array}{l}
{l_{nm}=a\qquad m=2,\ldots,M\enskip\mbox{and}\enskip
n=1,\ldots,N}\\
0\le l_{n1}\le+\infty\qquad n=1,\ldots,N\\
0\le\phi_{nm}\le 2\pi\qquad m=1,\ldots,M\enskip\mbox{and}\enskip
n=1,\ldots,N
\end{array}
\right.
\right\}
}{domaind}
At this point, using Eqs.~\ceq{fdfdf} and \ceq{jacobian} we go
back to 
cartesian coordinates:
\begin{eqnarray}
\Psi[J]&=&\lim_{N\to\infty}\lim_{M\to\infty}\int_{D}\prod_{n,m}d\mathbf R_{nm}
\int_{-\infty}^{+\infty}\prod_{n,m}d\bm\nu_{nm}\exp\left\{
-abc\sum_{n,m}\bm\nu^2_{nm}
\right\}
\nonumber\\ 
&
\times&
\prod_{n,m}\delta(\mathbf R_{nm}
-\bm\varphi_{\bm\nu,nm})\exp\left\{
ab\sum_{n,m}\mathbf J_{nm}\cdot\mathbf R_{nm}
\right\} \label{psijbeforelast}
\end{eqnarray}
The domain $D$ in cartesian coordinates is given by all $\mathbf
R_{nm}$'s in the two dimensional plane subjected to the constraints
\ceq{constao}: 
\beq{
D=\left\{
   \{
     \mathbf R_{nm}
   \}
   \left|
         \begin{array}{c}
               \mathbf R_{nm}\in \mathbb{R}^2\qquad\qquad m=1,
                \ldots,M\qquad
                n=1,\ldots,N\\ 
                \frac{
                |\mathbf R_{nm}-\mathbf R_{n(m-1)}|^2}
                     {a^2}=1\qquad m=2,\ldots,M\qquad 
                      n=1,\ldots,N 
         \end{array}
   \right.
\right\}
}{domaindcartcoord}
Finally, we rewrite the path integral in Eq.~\ceq{psijbeforelast} in
its continuous form. The result is:
\beq{
\Psi[J]=\int_{\mathbf R^{\prime\, 2}=1}{\cal
  D}\mathbf R \int{\cal D}\bm\nu e^{-c\intsym\bm\nu^2} 
\delta(\mathbf R -\bm\varphi_{\bm\nu})e^{\intsym \mathbf J\cdot
  \mathbf R}
}{psijaftercontlim}
The right hand side of the above equation
coincides exactly with the right hand side of
Eq.~\ceq{psitildejel}. This proves the equivalence between the
generating functional $\Psi[J]$ of the GNL$\sigma$M and the generating
functional $\tilde\Psi[J]$ of the stochastic process of
Eqs.~\ceq{langevequ}--\ceq{constrvarphinu}.

\section{Conclusions}
In this work it has been shown that the GNL$\sigma$M is related
to a stochastic process which, after discretization,
 describes the Brownian motion of $N$ beads subjected to the
 constraints \ceq{constao}. These constraints enforce the conditions
 that the 
 links connecting the beads are of fixed length. More in details, it
 has been proved that the generating functional 
$\Psi[J]$ of the GNL$\sigma$M coincides with the
generating functional $\tilde \Psi[J]$ of the solutions of the
Langevin equation 
\ceq{langevequ} and of the constraint \ceq{constrvarphinu}.
The fact that the two functionals are equal was not a priori obvious,
because they differ by the delta function $\delta(\mathbf R'^2-1)$ 
which
contains quadratic powers of the fields. 
If $\delta(g(\mathbf R))$ is a delta function imposing the condition
$g(\mathbf R)=0$, then in general the following identity is valid:
\beq{
\int {\cal D}\mathbf R
f(\mathbf R)\delta(g(\mathbf R))=\int_{g(\mathbf R)=0}{\cal
  D}\mathbf R
f(\mathbf
R){\det}^{-1}\left|
\frac{\delta{g(\mathbf R)}}{\delta \mathbf R}
\right| 
}{idefin}
If in our case the functional determinant appearing in the right hand
side of Eq.~\ceq{idefin} would be not trivial, then there would be no
chance that \ceq{psitildejel} and \ceq{beforedelta} coincide.
Luckily, it turns out that, after passing to the pseudo--polar coordinates
\ceq{pseudopolar}, the delta function  $\delta(\mathbf R'^2-1)$
produces
 just
a functional determinant which is a trivial constant.

\section{Acknowledgements}
This work has been financed
by the Polish Ministry of Science and Higher Education, scientific
project N202 156 
31/2933.  
F. Ferrari gratefully acknowledges also the support of the action
COST~P12 financed by the European Union and the hospitality of
C. Schick at the University of Rostock.
The authors  would like to thank
V. G. Rostiashvili and T. A. Vilgis for fruitful discussions.

\end{document}